\documentclass[12pt,preprint]{aastex}
\usepackage{epsfig}
\usepackage{rotating}
\usepackage{graphicx}
\def\xmm{{\sl XMM}-Newton}
\def\psr{PSR~B0950+08}
\def\psrr{PSR~B1929+10}
\def\fp{f_{\rm p}}
\def\ed{\dot{E}}
\def\lnon{L^{\rm nonth}}
\def\nh{n_{\rm H,20}}
\def\tbb{T^\infty_{\rm bb}}
\def\rbb{R^\infty_{\rm bb}}
\def\tpc{T_{\rm pc}}
\def\rpc{R_{\rm pc}}

\def\lbol{L_{\rm bol}}
\def\lbolpc{L_{\rm bol}^{\rm pc}}
\def\epc{\eta^{\rm pc}}
\def\enon{\eta^{\rm nonth}}

\newcommand{\gapr}{\raisebox{-.6ex}{\mbox{
$\stackrel{>}{\mbox{\scriptsize$\sim$}}\:$}}}
\newcommand{\lapr}{\raisebox{-.6ex}{\mbox{
$\stackrel{<}{\mbox{\scriptsize$\sim$}}\:$}}}

\begin{document}

\shortauthors{V.\ E. Zavlin and G.\ G. Pavlov}
\shorttitle{X-ray emission from \psr}
\title{X-ray emission from the old pulsar B0950+08}
\author{Vyacheslav E.\ Zavlin\altaffilmark{1}\altaffiltext{1}{
Observatoire Astronomique, 11 rue de l'Universit\'e, 
67000 Strasbourg, France} and
George G.\ Pavlov\altaffilmark{2}\altaffiltext{2}{
Dept.\ of Astronomy and Astrophysics, The Pennsylvania State
University, 525 Davey Lab, University Park, PA 16802}
}
\begin{abstract}
We present the timing and spectral analyses of the 
\xmm\ data on the 17-Myr-old, nearby radio pulsar B0950+08.
This observation revealed pulsations 
of the X-ray flux of \psr\ at its radio period, $P\simeq 253$ ms.
The pulse shape and pulsed fraction are apparently different 
at lower and higher energies of the observed 0.2--10 keV energy range, 
which suggests that the radiation
cannot be explained by a single emission mechanism.
The X-ray spectrum of the pulsar can be fitted with a power-law
model with a photon index $\Gamma=1.75\pm0.15$ 
and an (isotropic) luminosity
$L_X=(9.8\pm0.2)\times 10^{29}$ erg s$^{-1}$ 
in the 0.2--10 keV. Better fits are obtained with two-component, 
power-law plus thermal, models with 
$\Gamma=1.30\pm 0.10$ and $L_X=(9.7\pm 0.1)\times 10^{29}$ erg s$^{-1}$
for the power-law component that presumably originates from the pulsar's
magnetosphere. The thermal component, dominating at $E\lapr 0.7$ keV,
can be interpreted as radiation from heated polar caps on the neutron
star surface covered with a hydrogen atmosphere. The
inferred effective temperature, radius, and bolometric
luminosity of the polar caps are 
$\tpc\approx 1$ MK, $\rpc\approx 250$ m, 
and $L_{\rm pc}\approx 3\times 10^{29}$ erg s$^{-1}$. 
Optical through X-ray nonthermal spectrum of the pulsar
can be described as a single power-law with $\Gamma=1.3$--1.4
for the two-component X-ray fit.
The ratio of the nonthermal X-ray (1--10 keV) luminosity
to the nonthermal optical (4000--9000\,\AA)
luminosity, $\approx 360$, is within the range of $10^2$--$10^3$
observed for younger pulsars, which suggests that the magnetospheric
X-ray and optical emissions are powered by the same mechanism in
all pulsars, independent of age and spin-down power.
Assuming a standard neutron star radius,
the upper limit on the temperature of the bulk of
the neutron star surface, inferred from the optical
and X-ray data, is about 0.15 MK.
We also analyze X-ray observations of several other old pulsars, B2224+65,
J2043+2740, B0628--28, B1813--36, B1929+10, and B0823+26,
and compare their properties with the those of \psr.
\end{abstract}
\keywords{pulsars: individual (B0950+08, B2224+65, J2043+2740, B0628-28, B1813--36, B1929+10, B0823+26) --- stars: neutron --- X-rays: stars}
\section{Introduction}
Radiation from spin-powered pulsars at X-ray through optical wavelengths 
consists of thermal and nonthermal components
(see Kaspi, Roberts, \& Harding 2004 for a recent review).
In very young pulsars, 
with characteristic ages $\tau=P/(2\dot{P})\lapr 10$ kyr
($P$ and $\dot{P}$ are the pulsar's spin period and its derivative), 
such as the Crab pulsar, PSR B1509--58, and PSR B0540--69, 
thermal radiation from the neutron star (NS)
surface is buried under a powerful
nonthermal component from the NS magnetosphere,
whose spectrum can be described by a power-law (PL) model, sometimes
with a slowly varying slope.
The luminosity of the magnetospheric component decreases with pulsar's
age faster than the thermal luminosity (at $\tau\lapr 1$ Myr),
so that the thermal component, with
typical surface temperatures of $\sim 0.5$--1 MK,
dominates in the X-ray (and, sometimes, far-UV) emission from
middle-aged pulsars ($\tau\sim 0.01$--1 Myr; 
e.g., PSRs B0656+14, B1055--52, J0538+2817, Vela, and Geminga --- see 
Pavlov, Zavlin \& Sanwal 2002; Zavlin \& Pavlov 2004).
Old ($\tau\gapr 1$ Myr) NSs cool down very fast
(unless a heating process operates in the NS interiors), so that
the bulk of their surface radiation is emitted at optical-UV wavelengths.
However, as predicted by virtually all pulsar models,
small spots around the NS magnetic poles can be 
heated up to X-ray temperatures
by relativistic particles generated in the pulsar acceleration zones. 
These polar caps (PCs) are too small to be detectable in the optical,
but they can be seen in soft X-rays.
Whether the PC emission can be detected in X-rays, and the bulk surface
emission in the optical, depends on the intensity of the magnetospheric
component, which is is rather poorly investigated because old pulsars are
very faint objects beyond the radio band.
Thus, studying the X-ray and optical emission from old pulsars 
allows one to understand evolution of the magnetospheric radiation
at large ages and low spin-down energy losses,
constrain the pulsar models through observations of PCs,
and constrain the internal heating mechanisms through 
observations of thermal emission from the NS surface.

Similar problems have been addressed in X-ray and UV observations of
the nearest millisecond (recycled) pulsar J0437--4715 
($P\simeq 6$ ms, $\tau\simeq 5$ Gyr, $d\simeq 140$ pc).
Observations with {\sl ROSAT} and {\sl Chandra} have shown that its
PC emission dominates in soft X-rays, while the magnetospheric
radiation is responsible for the hard X-ray tail, at $E\gapr 3$ keV
(Zavlin \& Pavlov 1998; Zavlin et al.\ 2002).
Far-UV radiation from this pulsar shows a thermal spectrum,
emitted from the bulk of the NS surface with a temperature 
of $\sim 0.1$ MK (Kargaltsev, Pavlov, \& Romani 2004). 
However, millisecond pulsars represent a separate class
because they were spun up (recycled) by accretion in close binary systems.
Their spin-down power, $\ed$, is, as rule, much higher than that of
``ordinary'' old pulsars, which can affect their magnetospheric emission.
Heating mechanisms in these rapidly spinning objects
can also be quite different. Therefore, ordinary old pulsars deserve 
a separate investigation.

Only two ordinary old pulsars, B1929+10 ($P=0.226$ s, $\tau=3.1$ Myr, 
$\ed=3.9\times 10^{33}$ erg s$^{-1}$)
and B0950+08 ($P=0.253$ s, $\tau=17.4$ Myr, 
$\ed=5.6\times 10^{32}$ erg s$^{-1}$)
were detected outside the radio band in the pre-{\sl Chandra}/\xmm\ era.
For \psrr, 
{\sl ROSAT} and {\sl ASCA} observations provided some temporal and
spectral information (Yancopoulos, Hamilton \& Helfand 1994; Wang \&
Halpern 1997). Its X-ray light curve shows one broad pulse
per period, with a pulsed fraction $\fp\approx 35\%$.
The detected X-ray spectrum can be described by a PL
model with a photon index $\Gamma\approx 1.8$.
On the other hand, the same spectrum can be interpreted
as thermal emission with a blackbody (BB) 
temperature $\tbb\approx 0.5$ MK
and emitting area of a $\rbb\approx 600$ m radius.
The properties of the X-ray emission of \psr\
are even less certain. Although the pulsar was 
detected with {\sl ROSAT} (Manning \& Willmore 1994), 
the small number of source counts 
collected ($\sim 50$) precluded a meaningful data analysis.
\psr\ was also observed with {\sl ASCA}, but it was unresolved from
a brighter AGN. 

Both \psrr\ and \psr\ were detected in the UV-optical by
Pavlov, Stringfellow, \& C\'ordova (1996)
with the {\sl Hubble Space Telescope} ({\sl HST}) in a very broad band,
$\approx 2400$--4600\,\AA.
Further observations with narrower filters have shown that
their optical radiation is likely of a nonthermal origin
(Mignani et al.\ 2002; Zharikov et al.\ 2004).
In particular, the BVRI spectrum of \psr\ was found to be consistent
with a PL model with a broad range of photon
indices, $\Gamma_{\rm opt}=1.25$--2.05. Because of poor 
quality of the X-ray data available,
the connection of the optical and X-ray spectra remained unclear.

The high sensitivity and spectral resolution of {\sl Chandra} and \xmm\
have provided new opportunities to study old pulsars. In this paper
we present the analysis and interpretation of a recent \xmm\ 
observation of \psr, the oldest ordinary pulsar detected outside
the radio wavelengths.
We describe the \xmm\ data in \S\,2 and present the results of 
our timing and spectral analysis in \S\,3. 
Interpretations of the broad-band spectrum are discussed in \S\,4. 
In \S\,5 we compare the observed properties of \psr\ 
with the pulsar models and observations of other pulsars, 
including six old pulsars observed in X-rays.
We summarize our results in \S\,6.

\section{\xmm\ observation of \psr}
\xmm\ observed \psr\ on 2002 May 8--9 (orbit 442)
for 83.9 ks and 58.0 ks effective exposures with the EPIC\footnote{
European Photon Imaging Camera.}-MOS and EPIC-pn instruments,
respectively. 
For about 28 ks EPIC-MOS1 was operated in Full Frame mode 
providing an image of a large area, $r\sim 14'$, 
with time resolution of 2.6 s. For the rest of the time
it was in Timing mode, which provides high time resolution 
of 1.5 ms at the expense of one dimension of the image.
EPIC-MOS2 was in Full Frame mode for the whole exposure.
EPIC-pn was used in Small Window mode which covers a 
$4\farcm37\times 4\farcm37$ region and provides a 
$5.7$ ms time resolution.
All the EPIC instruments were used with Medium filters.
Two Reflection Grating Spectrometers (RGSs)
observed the target in the standard Spectroscopy mode for 84.4 ks.

Due to the high sensitivity and relatively high time resolution,
the data collected with EPIC-pn are most suitable
for studying the X-ray emission from \psr.
We used the EPIC-pn data processed with the most recent version
of the SAS package\footnote{{\tt http://xmm.vilspa.esa.es}}
(v.\,6.0.0).
Time intervals with strong background flares at the beginning and the
end of the observation were excluded, which, with account for the 70\%
efficiency of Small Window mode, resulted in a 51.1 ks effective exposure.
The pulsar is clearly detected in the EPIC-pn image at 
$\alpha_{\rm J2000}=09^{\rm h} 53^{\rm m} 9\fs34$ and
$\delta_{\rm J2000}=07^\circ 55' 35\farcs2$, 
which differs from the radio position (Table 1) by 
$1\farcs1$, well within the $2''$--$3''$ uncertainty of 
the EPIC absolute astrometry.
The shape of the EPIC-pn image of the pulsar is consistent with that
expected for a point-like source. We put a 3$\sigma$ upper limit of
$<2\times 10^{-14}$ photons s$^{-1}$ cm$^{-2}$ arcmin$^{-2}$, on the 
surface brightness of emission from a possible pulsar-wind nebula
at distances $\sim 1'$--$2'$ from the pulsar.

We extracted the source (plus background)
photons from a $30''$-radius circle centered at the pulsar position,
which encircles about 82\% of all detected
source counts\footnote{
See the \xmm\ Users' Handbook at {\tt http://xmm.vilspa.esa.es}.}.
For the analysis, we selected only single and double events 
(with photon-induced charge detected in a single CCD pixel and two 
adjacent pixels with horizontal or vertical splits)
for which the instrumental response is known (those events 
provide about 98\% of all detected photons).
The extracted count rate spectrum 
shows strong excesses of single and double events
at energies below 0.2 and 0.4 keV, respectively
(cf.\ van Kerkwijk et al.\ 2004), probably caused by
the internal EPIC-pn `flickering'. Because the distribution of these  
detector background events is rather nonuniform over
the field of view, they cannot be reliably subtracted; 
therefore we had to discard single events with $E<0.2$ keV 
and double events with $E<0.4$ keV. The background was 
estimated from several
circles of the same $30''$-radius centered at different positions.
After the correction for the 82\% encircled count
fraction, we found the source count rate of $0.033\pm 0.001$ 
counts s$^{-1}$ for single events in the 0.2--10 keV range.
Adding doubles with $E>0.4$ keV increases this value by 26\%. 

Because of the much lower sensitivity, 
the EPIC-MOS instruments detected the pulsar
at a very low rate of about 0.007 counts s$^{-1}$, 
which makes the EPIC-MOS1 data taken in Timing mode
hardly useful for the analysis 
because of a strong background contamination (the same concerns the RGS
data where the source count rate is even smaller).
We also do not use the EPIC-MOS data taken in the imaging mode because of
uncalibrated changes in the low-energy spectral response of the
EPIC-MOS detectors\footnote{See 
{\tt http://www.src.le.ac.uk/projects/xmm/technical/tuebingen0203.html}.}.

\section{Data analysis}
\subsection{Timing}
For the pulsation search, we used 1,568 
events (singles and doubles) within the $r=30''$ circle,
in the 0.2--5.0 keV range,
where the pulsar radiation prevails over the background.
The source was estimated to contribute about 84\% of the selected events.
The photon times of arrival were transformed to 
the Solar System Barycenter with the ``barycen'' task of the SAS package.
Using the TEMPO code\footnote{A program for the analysis of pulsar
timing data maintained and distributed by Princeton University and the
Australia Telescope National Facility. It is available at
{\tt http://www.atnf.csiro.au/research/pulsar/timing/tempo/}}, 
we calculated the radio ephemeris (see Table 1) based on
observations of \psr\ in 2002 
at the Torun Radio Astronomy Observatory (the data
provided by Wojciech Lewandowski).
For the radio-pulsar frequency at the epoch of the \xmm\ observation, 
the $Z^2_n$ test (e.g., Buccheri et al.\ 1983) yields values 
$Z^2_1=30.7$, $Z^2_2=61.4$, and $Z^2_3=72.7$,
of which the latter is most significant: 
the probability to obtain this
value by chance in one trial is $1.1\times 10^{-13}$.

Using the radio data and the TEMPO code,
we extracted the X-ray pulse profiles in radio  phases ($\phi=0$
corresponds to the phase of radio pulse maximum)
from the single-event data in four energy ranges, 0.2--0.5,
0.5--1, 1--5  and 0.2--5 keV (see Fig.\ 1). At energies $E>0.5$ keV
(which contribute the bulk of the total detected emission) the 
light curves show two distinct peaks, centered at phases 
$\phi\approx 0.25$ and 0.85.
The shapes of these peaks and the large pulsed fraction
suggest a predominantly nonthermal 
origin of the pulsar emission at $E>0.5$ keV.
Interestingly, the separation $\Delta\phi \approx 0.4$ between the
peaks is about the same as $\Delta\phi = 0.42$ between the radio 
pulse and interpulse, albeit the radio interpulse is a factor of 50
weaker than the main pulse (Everett \& Weisberg 2001).

The shape of the light curve at lower energies looks different. 
The highest peak in the light curves at $E>0.5$ keV 
(at $\phi\approx 0.25$)
is strongly suppressed in the 0.2--0.5 keV pulse profile, so that
the peak at $\phi\approx 0.85$ becomes dominant in this energy band. 
Additionally, the pulsed fraction in the 0.2--0.5 keV band is reduced 
to about 33\% (albeit with large uncertainties). 
To check whether the change in the shapes of the pulse profiles 
shown in the upper three panels
of Figure 1 is statistically significant, we applied the $\chi^2$
test as described by Zavlin \& Pavlov (1998).
This test gives a probability of 97.1\% 
that the shapes of the light curves extracted in the 0.2--0.5 keV
and 1--5 keV bands are different.
The same test shows that the shapes of 
the 0.2--0.5 keV and 0.5--1 keV light curves
are different with a probability of 94.7\%.
Therefore, it seems quite plausible that the change of the 
pulse shape and pulsed fraction is real.
Since one can hardly expect significant changes of pulsations in
a single nonthermal component over such a narrow energy range, 
we consider the apparent energy dependence of pulsations as
an indication of the presence of another emission
mechanism(s), with the main contribution at $E\lapr 0.5$ keV.

\subsection{Spectral analysis of the phase-integrated spectrum}
The EPIC-pn data on \psr, reduced as described in \S\,2, were used for 
the spectral analysis. 
Because of the small number of counts collected, phase-resolved
spectroscopy cannot produce useful results, so we have to use
the phase-integrated spectrum. The EPIC-pn spectrum of
the pulsar was binned in 32 spectral bins in the 0.2--10 keV
range with at least 40 source counts per bin.
This provides the signal-to-noise ratio $S/N=5$--6
in the spectral bins below 2 keV and $S/N=3.5$--4 at $E>2$ keV.
We checked that varying the number of source counts per bin
between 30 and 50 (with $S/N>3$ in each of the spectral bins) 
makes no significant effect on the results presented below.
We also verified that
the 0.4--10 keV spectrum composed of double events is well
consistent with that obtained from single events.
Since the double-event spectrum is more noisy 
and contains only a quarter of the whole flux, we used only
the single-event data for the further spectral analysis.

We started from fitting the pulsar count rate spectrum with
one-component thermal models (blackbody [BB] and NS atmosphere)
and found the best-fit models to be
much softer than the data at $E\gapr 2$ keV,
which resulted in an unacceptably large minimum $\chi_\nu^2$ ($> 6.5$
for $\nu=29$ degrees of freedom [d.o.f.]). 
A fit with two thermal components (e.g., BB+BB) 
still has the same problem at $E\gapr 5$ keV. 
In addition, it yields an unreasonably high temperature,
$T\gapr 10$ MK, and an unrealistically small size ($\lapr 2$ m radius)
for the harder component.
This means that the observed spectrum at higher energies
is of a nonthermal (presumably magnetospheric)
origin, in agreement with the timing results (\S\,3.1). 

The (absorbed) PL model fits the data much better
($\chi^2_\nu=1.42$ for 29 d.o.f. --- see Fig.\ 2)\footnote{
Systematic residuals at $E\lapr 0.5$ keV can be attributed to
uncertainties in the calibration of the EPIC-pn detector.},
yielding the photon index $\Gamma=1.75\pm 0.15$ 
and the absorbing hydrogen column density 
$\nh\equiv n_{\rm H}/(10^{20}\,{\rm cm}^{-2})=2.9\pm 2.1$ ($1\,\sigma$
errors are given here and below).
The observed (absorbed) flux in the 0.2--10 keV range, 
$F_X=(1.1\pm 0.1)\times 10^{-13}$ erg s$^{-1}$ cm$^{-2}$,
corresponds to the (unabsorbed) isotropic
luminosity\footnote{Here and below
all the luminosities, efficiencies, and radii are given for a distance
$d=262$ pc (Brisken et al.\ 2002).}  
$\lnon_{\rm 2-10\,keV}=(9.8\pm 0.2)\times 10^{29}$ erg s$^{-1}$
and the ``nonthermal X-ray efficiency'' $\enon_{\rm 2-10\,keV}
\equiv \lnon_{\rm 2-10\,keV}/\ed = (1.75\pm 0.04)\times 10^{-3}$.
Although the PL fit is (marginally) acceptable, 
and the derived fitting parameters look reasonable, the fit 
residuals show that the model is 
somewhat softer than the data at $E\gapr 5$ keV (Fig.\ 2). 
This can be considered as an additional
argument supporting the suggestion in \S\,3.1
that the X-ray emission of \psr\ consists of at least two 
components of different origin.

It is natural to assume (see \S\,1) that, in addition to the
magnetospheric emission, thermal emission from heated PCs
contributes to the observed X-ray flux (see \S\,1).
As a first step in the two-component, nonthermal plus thermal,
fit, we added a standard BB model to the PL component.
The fit gives the BB temperature $\tbb=1.75\pm 0.22$ MK,
effective radius $\rbb=50\pm 32$ m,
and bolometric luminosity
$L^\infty_{\rm bb}=(1.6\pm 0.8)\times 10^{29}$ erg s$^{-1}$.
(The superscript $^\infty$ is used for the values as measured 
by a distant observer. The values as measured at the NS surface are 
$T_{\rm bb}=g_r^{-1} \tbb$, $R_{\rm bb}= g_r \rbb$, and
$\lbolpc=g_r^{-2} L^\infty_{\rm bb}$,
where $g_r=[1+z]^{-1}=[1-2GM/c^2R]^{1/2}$.)
The PL photon index of the nonthermal component is
$\Gamma=1.31\pm 0.14$,
while the nonthermal luminosity,
$\lnon_{0.2-10\,{\rm keV}}=(9.7\pm 0.1)\times 10^{29}$ erg s$^{-1}$,
corresponds to the nonthermal X-ray efficiency
$\enon_{0.2-10\,{\rm keV}}=(1.73\pm0.02)\times 10^{-3}$.
The hydrogen column density derived from the BB+PL fit,
$\nh=3.3\pm2.2$, is almost the same as obtained from the PL fit above,
and it does not contradict to independent estimates\footnote{
The total Galactic column density of neutral hydrogen
along the direction to \psr\ is $n_{\rm HI,20}\approx 3$
({\tt http://heasarc.gsfc.nasa.gov/cgi-bin/Tools/w3nh/w3nh.pl}).
UV observations of stars at distances 110--160 pc from the pulsar
give $\nh$ in a range of 0.1--10, with large uncertainties
({\tt http://archive.stsci.edu/cgi-bin/ismform}).}.
The quality of the BB+PL fit, $\chi^2_\nu=1.20$ for $\nu=27$,
is somewhat better than that obtained with the single PL model,
although, according to the standard {\sl F}-test, 
the probability that the spectral data require the second
component is not very high, 96.1\%.

The radius obtained from the BB+PL fit is much smaller than the expected
PC radius, $\sim R (2\pi R/cP)^{1/2} = 288\, (R/10\,{\rm km})^{3/2}$ m.
In addition, the isotropic BB emission from two symmetric PCs
cannot reproduce the strong pulsations for realistic $M/R$ ratio
(e.g., Zavlin, Shibanov \& Pavlov 1995).
However, the spectrum and the beaming of thermal emission can be
substantially different from those of BB emission if the NS surface
is covered by a light-element (hydrogen or helium) atmosphere
(see Zavlin \& Pavlov 2002 for a review on NS atmospheres).
For example, if there is as few as $10^{13}$ g of hydrogen in the NS
surface layers (e.g., as a result of accretion from 
the interstellar medium 
and/or fallback of a fraction of the envelope ejected
during the supernova explosion), this hydrogen will rise to the surface
forming an optically thick, purely hydrogen atmosphere.
The spectrum emerging from a hydrogen atmosphere,
which is almost fully ionized at the expected temperatures $\sim 10^6$ K
and magnetic fields $\sim 10^{11}$--$10^{12}$ G, is 
considerably harder at higher energies
than the Planck spectrum with the same effective temperature,
so that fitting the observed spectrum with a BB model gives 
a higher temperature and a smaller size, compared to their actual values
(e.g., Pavlov et al.\ 1995). In addition,
strong magnetic fields in NS atmospheres
affect the beaming and the spectrum of the emergent radiation; 
in particular, the spectrum from a strongly ionized atmosphere shows
an absorption line at the electron cyclotron energy 
$E_{\rm c}=1.16\,(B/10^{11}\,{\rm G})\, g_{\rm r}$ keV.

To determine the effective temperature $\tpc$
and the radius $\rpc$ of PCs under the 
assumption of hydrogen atmosphere, we assume that the thermal X-rays are
emitted from two similar, uniformly heated PCs around 
the magnetic poles of a centered magnetic dipole. 
In addition to the temperature and radius,
the (phase-integrated) model spectrum depends
on the magnetic field, inclination $\zeta$ of the rotation axis to 
the line of sight, magnetic inclination $\alpha$ between the 
magnetic and rotational axes, and the mass-to-radius ratio, $M/R$, 
that is responsible for the redshift
and bending of photon trajectories in the strong 
gravitational field of the NS.
Since a fit with so many fitting parameters would be 
essentially unconstrained, we have to choose fixed values for 
some of them. 
Based on the results of the single PL and BB+PL fits,
we restricted the absorption column density in a plausible range,
$2<\nh < 4$, chose, as an example, $B=5\times 10^{11}$ G 
(which is the magnetic field at the poles of an orthogonal rotator
in the idealized magnetic dipole radiation model,
$B_{\rm p}=6.4\times 10^{19} [P\dot{P}]^{1/2}$ G),
$\zeta =\alpha = 45^\circ$, $M=1.4 M_\odot$, and $R=10$ km
($g_r=0.77$, $z=0.30$).
We calculated the model specific intensities of emitted radiation
on a $\tpc$ grid and incorporated them in the XSPEC package.
These intensities provide the spectra at different rotational
phases as well as the phase-integrated spectrum 
(see Zavlin et al.\ 1995 and Zavlin \& Pavlov 1998 for more detail).
The two-component (PC+PL) fit ($\chi_\nu^2=1.17$, for 27 d.o.f.)
resulted in the PC (effective) temperature $\tpc=1.06\pm 0.16$ MK 
and radius $\rpc=250\pm 30$ m (as measured at the NS surface),
the PL photon index $\Gamma=1.29\pm 0.12$,
and $\nh = 3.2\pm 0.8$.
Varying $\zeta$ and $\alpha$ between $0^\circ$ and 
$90^\circ$ and the mass-to-radius ratio in a plausible range of
$M/R=(1.0$--$1.8)\,M_\odot/(10\,{\rm km})$
changes the best-fit temperature by $\lapr 0.1$ MK and the best-fit
radius by $\lapr 90$ m.
The bolometric luminosity of the thermal component
is $\lbolpc=(2.7\pm 0.8)\times 10^{29}$ erg s$^{-1}$ (for two PCs),
while the nonthermal component 
has the same luminosity as the best BB+PL fit.
The best PC+PL fit, together
with the two spectral components, is shown in Figure 3.
As seen from this plot, the thermal component dominates at $E<0.7$ keV,
providing about 60\% of the total flux observed in the 0.2--0.5 keV range.
We checked that varying the magnetic field makes almost no effect on
the inferred PC parameters as far as the cyclotron energy 
remains in the Wien tail, $B\gapr 2\times 10^{11}$ G.

\subsection{Thermal light curve analysis}
The PC model used in \S\,3.2 for
the analysis of the phase-integrated spectrum predicts
a large variety in pulse profiles of the thermal emission,
depending on the pulsar geometry, mass-to-radius ratio, and surface
magnetic field (Zavlin et al.\ 1995). 
Therefore, these parameters can be constrained from the comparison of the
model light curves with the observed ones (cf.\ Pavlov \& Zavlin 1997).
To make such a comparison for \psr, one first needs to 
single out the contribution of the possible thermal component
in the observed light curve.
Assuming that the pulse shape of the nonthermal flux does not 
depend on photon energy in the observed X-ray range,
we scaled the nonthermal light curve of the 1--5 keV band (where the PC
contribution is virtually negligible, $\lapr 5\%$) to the 0.2--0.5 keV
band, chosen for the thermal light curve analysis, and subtracted
the scaled nonthermal light curve.
The scaling factor of 0.31 was estimated as
the ratio of the number of counts provided by the nonthermal 
component of the best PC+PL fit of \S\,3.2 in the 0.2--0.5 kev band to 
the total number of counts in the 1--5 keV band
(varying the parameters of the PL component within the 1$\sigma$
confidence range changes this factor by 15\% only,
which makes a negligible effect on the results presented below). 
The result of the light curve subtraction is shown in Figure 4.
Although the residual pulse profile is noisy
because of the scanty statistics, it 
shows a structure around the phase $\phi=0.85$ 
which can be interpreted as a pulse of thermal emission.
The pulsed fraction of this light curve is fairly
large, with a lower limit of 35\%. 
Such a high pulsed fraction cannot be 
obtained with the isotropic BB radiation for any $\zeta$ and $\alpha$
unless the NS mass-to-radius ratio is so (unrealistically) small,
$(M/M_\odot)/(R/10\,{\rm km}) \lapr 0.25$ 
(i.e., gravitational redshift $z\lapr 0.04$),
that the effect of the gravitational bending is negligible.
Therefore, a strong magnetic field, $B\gapr 10^{11}$ G, is required
to explain the large pulsed fraction. In strong fields, such that 
$E_{\rm c}\gg E$, the angular distribution of the emergent 
radiation shows a narrow beam along the direction of the magnetic field
(where the atmospheric plasma is most transparent for radiation) and a 
rather broad hollow-cone beam around the magnetic field direction
(Pavlov et al.\ 1994; Zavlin \& Pavlov 2002).
If $\zeta$ and $\alpha$ are such that the narrow beam passes through 
the line of sight in the course of pulsar rotation, 
the observed flux can be very strongly modulated.

Because of the large errors in the residual light curve
shown in Figure 4, fitting it with
the multi-parameter PC models is hardly warranted. 
Instead, we crudely determined domains of parameters where this 
light curve qualitatively agrees with the models.
For several sets of $B$, $\zeta$, $\alpha$, and $M/R$,
we found best-fit $\tpc$, 
$\rpc$, and $n_{\rm H}$ from spectral fits (as described in \S\,3.2),
used these values
to compute the model fluxes as a function of phase, and folded these
fluxes with the EPIC-pn response.
Comparing the model light curves with the residual light curve,
we found that a reasonably good agreement can be achieved, for example,
in the following ranges of parameters: $|\zeta - \alpha|\lapr 20^\circ$,
$30^\circ\lapr\zeta\lapr 75^\circ$, $B\approx (2$--$3)\times 10^{11}$ G,
and $(M/M_\odot)/(R/10\, {\rm km}) \approx 1.2$--1.6.
At larger $|\zeta -\alpha|$, the peak direction of the narrow beam 
(along the magnetic axis) remains too far from the line of sight,
which results in low modulation of the model flux.
The modulation is also too low at small $\zeta$ and/or $\alpha$,
while at $\zeta$ and $\alpha$ close to $90^\circ$ we would see two
peaks of comparable strength, 
in contradiction with the residual light curve.
At higher magnetic fields, the central beam becomes too narrow, so that
the pulse profiles with a sufficiently large pulsed fraction
are sharper than observed. 
An example of a reasonably good agreement is shown in
the upper panel of Figure 4 that presents a model light curve 
with $B=2\times 10^{11}$ G, $\zeta=35^\circ$, $\alpha=30^\circ$ 
and $[M/M_\odot]/[R/10\,{\rm km}]=1.4$ 
($\tpc=1.02$ MK, $\rpc=280$ m, and $\nh=3.0$). 
The model pulsed fraction is $\fp=68\%$.

In principle, independent estimates on the angles $\zeta$ and $\alpha$
can be obtained from the phase dependence of the position angle
of the linearly polarized radio emission (Radhakrishnan \& Cooke 1969).
For \psr, the estimates obtained by different authors
are summarized by Everett \& Weisberg (2001). Most of them 
suggest that \psr\ is an almost aligned rotator with only one pole
visible:  $\alpha\lapr 12^\circ$, $\zeta\lapr 5^\circ$.
At such small angles, the pulsed fraction of the
PC X-ray radiation, $\fp\lapr 15\%$, is too low
to be consistent with the residual light curve,
even with account for the uncertainty in the light curve subtraction.
The result of Everett \& Weisberg (2001),
$\zeta\simeq 53^\circ$ and $\alpha\simeq 75^\circ$
(an almost orthogonal rotator, with two poles visible),
differs significantly from the others. The pulse profile 
for the PC model with these angles and 
$B=2\times 10^{11}$ G, $M=1.4 M_\odot$ and $R=10$ km 
($\tpc=1.05$ MK and $\rpc=300$ m, $\nh=3.0$) is shown
in the lower panel of Figure 4. Its pulsed fraction is
$\fp=54\%$. Although the agreement does not look as good as 
in the upper panel, the deviation from the observational light curve 
is well within the uncertainties.

\section{X-rays versus optical}
The optical-UV fluxes measured in the 
{\sl HST}, {\sl VLT} and {\sl Subaru} observations of \psr\ 
(see Fig.\ 5) can be crudely described by a PL spectral 
model, with a photon index $\Gamma_{\rm opt}=1.65\pm 0.40$
(Zharikov et al.\ 2004)\footnote{The value for $\Gamma_{\rm opt}$
was obtained neglecting the interstellar extinction. 
The large scatter of fluxes in the different
optical bands and the resulting uncertainty of spectral slope
are likely due to errors in subtraction of the flux of 
background objects near \psr.}.
This means that both optical ($E\approx 1.5$--5 eV) 
and X-ray ($E\gapr 1$ keV) emission from \psr\ are 
nonthermal, most likely emitted by relativistic
particles in the pulsar magnetosphere. From observations of younger 
pulsars, we know that the optical and X-ray parts of the nonthermal 
spectrum can be described by the same PL model for some of them (e.~g.,
PSR B0656+14; see Pavlov et al.\ 2002),
while in others the optical spectrum is harder 
than the X-ray spectrum (i.e., $\Gamma_{\rm opt} < \Gamma_{\rm X}$), 
and it lies below the extrapolation
of the X-ray power law as, for example, in the Crab
pulsar (Kuiper et al.\ 2001).
The change of the spectral slope in the latter case can be 
attributed to, e.g., deviation of the spectrum of 
emitting electrons from the pure PL 
or, less likely, to different populations of electrons emitting in the 
optical and X-ray bands. 

The two possible interpretations of the X-ray spectrum of \psr\ 
(pure PL and PC+PL) correspond to the different 
broad-band (optical through X-rays) spectra described above.
The extrapolation of the single PL model with $\Gamma=1.6$--1.9 
towards lower energies exceeds the observed optical spectrum by
a factor of 10--30, depending on $\Gamma$ and optical frequency.
In this case, we have to assume
a change in the PL slope somewhere in the EUV or soft-X-ray range
and a rather small $\Gamma_{\rm opt}\lapr 1.2$,
as demonstrated in the upper panel of Figure 5.
On the contrary, the extrapolation of the PL component of the PC+PL
model, which we consider more plausible than the single PL model, 
goes through the optical
points, so that the nonthermal spectrum in the 1 eV -- 10 keV range
can be described by a single PL with $\Gamma=1.3$--1.4.
The lower panel of Figure 5 shows the broad-band spectrum for
$\Gamma=1.35$; the corresponding broad-band nonthermal luminosity 
is $\lnon_{\rm 1\,eV - 10\,keV} = 1.1\times 10^{30}$ erg s$^{-1}$ = 
$1.8\times 10^{-3}\ed$,
of which the optical luminosity is a small fraction 
(e.g., $L_{1-5\, {\rm eV}}=5.2\times 10^{27}$ erg s$^{-1}$ = 
$4.7\times 10^{-3} \lnon_{\rm 1\,eV - 10\,keV}$).

Because of the small emitting
area, the thermal PC radiation falls 
about 3 orders of magnitude below the detected optical fluxes.
On the other hand, one cannot rule out the possibility that the
bulk of the NS surface is hot enough to emit 
detectable thermal radiation. Assuming a constant slope of the
optical through X-ray spectrum,  
we estimate an upper limit on the NS 
brightness temperature in the optical-UV: $T^\infty \lapr 0.15$ MK,
for $R^\infty = 13$ km and a plausible optical extinction, 
$E(B-V)= 0.05$. This limit is much higher than the expected
surface temperature of a 20-Myr old, passively cooling NS
(Yakovlev \& Pethick 2004).
However, various processes of NS heating 
(see Schaab et al.\ 1999 for a review)
can heat the surface up to even higher temperatures at this NS age.
Kargaltsev et al.\ (2004) detected thermal emission from the much older
($\tau = 4.9$ Gyr) millisecond pulsar J0437--4715, corresponding to the
surface temperature of about 0.1 MK. Therefore, one could expect 
at least a similar
surface temperature for the younger \psr, unless millisecond pulsars are
heated by a specific mechanism that does not work in ``ordinary'' pulsars
(e.g., Reisenegger 1997). To illustrate the effect of such temperatures
on the broad-band spectrum of \psr, we assumed a bulk surface temperature
$T_s^\infty = 0.1$ MK 
and plotted the corresponding (BB) spectrum in the
lower panel of Figure 5, for a NS radius $R^\infty=13$ km.
This thermal component (TS) gives virtually no contribution into
the observed X-ray range, but it would be observable in the far-UV range, 
at $\lambda \lapr 2000$ \AA.
 
\section{Discussion}
\subsection{Nonthermal emission from PSR B0950+08 and other pulsars}
Since the bulk of the X-ray and optical emission observed from \psr\
is nonthermal, it is interesting to compare it with the nonthermal
emission from other spin-powered pulsars.
We know from X-ray observations that the pulsar nonthermal luminosity 
shows a correlation with the spin-down power $\ed$.
For instance, Possenti et al.\ (2002) found a fit
$L_{\rm 2-10\,keV}^{\rm nonth}=
3.6\times 10^{31}(\ed/10^{35}\,{\rm erg\,s}^{-1})^{1.34}$ erg s$^{-1}$,
based on observations of 37 pulsars 
(mostly with {\sl ROSAT}, {\sl ASCA}, and {\sl Beppo-SAX}).
However, the observed luminosities showed large deviations, 
up to 2 orders of magnitude, from this dependence. 
The scatter becomes even larger if one includes
the upper limits for pulsars that were not detected in deep observations
(see some examples in \S\,5.3). 
Such a scatter can be naturally explained by
geometrical factors. Firstly, the actual luminosity can be substantially
different from the ``isotropic luminosity'' 
$4\pi d^2 F_X^{\rm unabs}$ because pulsar's
X-ray radiation is anisotropic, as we see from their pulsations, and even
powerful and radio-bright pulsars can be quite faint in X-rays if the 
rotating X-ray beam 
misses the Earth. Secondly, the properties of the pulsar's
magnetosphere and its radiation can strongly 
depend on the magnetic inclination
$\alpha$ that is unknown for most pulsars. 
Therefore, more physical is the 
``line of maximum luminosity'' in the $\ed$-$L_X^{\rm nonth}$ plane, 
such that all the observed luminosities
lie below that line. The ratio $\enon_{\rm max}\equiv 
L_{\rm max}^{\rm nonth}/\ed$ characterizes maximum efficiency of 
conversion of the spin-down power into X-ray radiation.  
Possenti et al. (2002) suggested 
$\enon_{\rm max}=0.02\, (\ed/10^{35}\,{\rm erg\, s}^{-1})^{0.48}$,
in the 2--10 keV band, which implies
that pulsars with low $\ed$ are less efficient X-ray emitters.

The derived luminosity of the nonthermal component
of \psr\ in the 2--10 keV range
is $\lnon_{\rm 2-10\,keV}\simeq 7.3\times 10^{29}$ erg s$^{-1}$, 
corresponding to $\enon_{\rm 2-10\,keV}=0.0013$.
These values exceed by a factor of 20 
the prediction of the $\lnon_X$--$\ed$ 
correlation derived by Possenti et al.\ (2002),
being close to the maximum values,
$L_{\rm max}^{\rm nonth} = 9.3\times 10^{29}$ erg s$^{-1}$ 
and $\enon_{\rm max}=0.0017$.

It is worthwhile to note that the nonthermal X-ray luminosity of \psr\
is only a factor of 2 lower than that of the much more powerful
($\ed = 3.8\times 10^{34}$ erg s$^{-1}$)
middle-aged ($\tau = 110$ kyr) pulsar B0656+14 (Pavlov et al.\ 2002),
which means that B0656+14 is a factor of 30--40 less efficient nonthermal
X-ray emitter than B0950+08.
Even less efficient, by a factor of about 600 with respect to B0950+08, 
is the younger ($\tau=11$ kyr) Vela pulsar (Pavlov et al.\ 2001). 
Based on these examples, it might be tempting to speculate that the
nonthermal X-ray efficiency grows with decreasing 
$\ed$ or increasing $\tau$.
However, there are many other middle-aged pulsars whose efficiency is
much higher than that of B0656+14 or Vela, even with account for larger
uncertainties in their distances. Therefore,
the most plausible reason for the high nonthermal 
X-ray efficiency of B0950+08
is not its low spin-down power or old age, but the above-mentioned
geometrical effects.

The optical part of the nonthermal spectrum of \psr\ also
shows an optical efficiency, 
$\eta_{\rm opt}\equiv L_{\rm opt}^{\rm nonth}/\ed$,
much higher than those of the optically detected middle-aged
pulsars (Geminga, B0656+14, and Vela), 
but comparable to those of most powerful, very young pulsars
(B0540--69 and Crab). Zharikov et al.\ (2004) argue that this shows
``non-monotonic evolution'' of the optical efficiency and speculate
about its physical explanations. However, 
the $\ed$ dependence of 
$L_{\rm opt}^{\rm nonth}/\lnon_{X}$ 
(where $L_{\rm opt}^{\rm nonth}$ is the nonthermal 
optical luminosity in the 4000--9000 \AA\ band
and $\lnon_{X}$ is that in the 1--10 keV range),
plotted in Figure 6 for a sample of 7 pulsars
(see also Table 2) shows that this ratio remains 
approximately constant, $\sim 0.003$ (with a factor of 3 scatter).
This means that a more plausible explanation is
the same geometrical effect that is responsible for 
the apparent growth of
the X-ray efficiency with age (at $\tau\gapr 10$ kyr), 
derived from the very limited sample.
This conclusion is further supported by the fact that some old, 
low-$\ed$ pulsars show quite low X-ray efficiencies, 
contrary to B0950+08 and B1929+10 (see \S\,5.3).
The approximately constant value of 
$L_{\rm opt}^{\rm nonth}/\lnon_{X}$
suggests a common origin of the magnetospheric
X-ray and optical emission and implies similar X-ray and optical beamings.

Finally, we can use the X-ray data to speculate about $\gamma$-ray
emission of \psr. First, we can estimate the maximum photon energy
$E_{\rm max}$
up to which the spectral slope can remain the same as in the optical
through X-ray range, from the condition that the total 
photon luminosity $\lnon$ does not exceed $\ed$. For example, it gives 
$E_{\rm max}< 140$ MeV for the PL model with $\Gamma = 1.35$
discussed in \S~4. Next, 
based on {\sl CGRO} EGRET observations,
Thompson et al.\ (1994) put an upper limit on the photon flux of \psr, 
$f_\gamma <5.2\times 10^{-7}$ photons s$^{-1}$ cm$^{-2}$
in the 100 MeV -- 1 GeV range, a factor of 7 higher than the
maximum possible contribution from the PL model of $\Gamma = 1.35$
into this energy range.
If, in future observations, $\gamma$-ray emission is 
detected at $E\gapr 140$ MeV or the detected $\gamma$-ray flux exceeds
the extrapolation of the X-ray PL spectrum
at $E\lapr 140$ MeV, it will mean that these $\gamma$-rays are generated
by a mechanism different from that responsible for the X-ray and optical 
emission.

\subsection{Properties of polar caps}
Comparison of parameters of heated PCs with 
the predictions of pulsar models
allows one to test these models and better understand the complex 
physical processes operating in spin-powered pulsars.
The PC temperature and radius 
inferred from the data analysis strongly depend on the assumed model  
for thermal emission, while the bolometric luminosity is less
sensitive to this assumption. 
For \psr, the harder NS atmospheric emission yields a factor 
of 2 lower surface temperature and a factor of 25 larger emitting 
area than those obtained with the BB model, whereas the bolometric
luminosities are approximately the same, 
$\lbolpc \approx 3\times 10^{29}$ erg s$^{-1}$, 
corresponding to the ``PC efficiency'' $\epc\equiv
\lbolpc/(2 \ed) \approx 2.4\times 10^{-4}$ (per one polar cap).
As we have shown in \S\,3.2, the assumption than the NS surface 
is covered with hydrogen leads to a PC radius, $\rpc=160$--340 m 
(depending on the pulsar geometry and mass-to-radius ratio),
that is in excellent agreement with the conventional 
estimate, $\rpc^*=(2\pi R^3/cP)^{1/2}=288$ m (for $R=10$ km).

Different pulsar models predict quite different PC luminosities
and efficiencies. A detailed analysis of PC heating,
relevant to old pulsars, has been
recently presented by Harding \& Muslimov (2001, 2002;
HM01 and HM02 hereafter). They updated 
the space charge limited flow
model (Arons 1981) by including the effects of general relativity
on the electric field induced above the NS surface
and taking into account inverse Compton scattering (ICS)
of thermal X-rays from the NS surface by primary particles (electrons)
accelerated in the NS magnetosphere.  In these models, the number of 
positrons accelerated
downward to heat the PC is a tiny fraction of the number of 
primary electrons that are accelerated upward.
The electron-positron pairs are created in the vicinity of 
NS surface in cascade processes initiated
by $\gamma$-ray photons emitted by the primary electrons.
HM01 and HM02 calculated the luminosity of a PC 
heated by positrons produced
through curvature radiation (CR) and radiation from ICS.
The former process dominates
in pulsars with higher magnetic fields and shorter periods.
Pulsars with lower fields and longer periods are not capable of 
producing pairs through CR so that the positrons
are produced through less efficient ICS,
which results in a lower PC luminosity. 

The period and magnetic field of \psr\
are very close to the boundary between these two regimes
(the death line for CR pair production;
see Fig.\ 1 of HM02), so both heating mechanisms should 
be considered. The PC efficiency and the luminosity of 
a single PC heated by particles created through CR
can be estimated from Figure 7 of HM01 or Figure 9 of HM02:
$\epc_{\rm CR}\approx 10^{-2}$ and $L^{\rm pc}_{\rm CR}
\approx 6\times 10^{30}$ erg s$^{-1}$,
exceeding by a factor of $\sim$40 the values we obtained
from the spectral analysis. 
On the other hand, heating through ICS (at $T_{\rm pc}\sim 1$ MK)
gives $\epc_{\rm ICS}\approx 10^{-5}$ and
$L^{\rm pc}_{\rm ICS} \approx 6\times 10^{27}$ erg s$^{-1}$
(see Figs.\ 8 and 9 of HM02), a factor of $\sim$24 lower
than the observational result. This means that either
PC heating by ICS-produced positrons is more efficient than calculated
by HM02 or, more likely, heating by CR-produced positrons 
is less efficient
in the transition regime near the death line for CR pair production.
The CR efficiencies were calculated by HM01 assuming 
a magnetic inclination angle
$\alpha = 60^\circ$; they are expected to decrease with
increasing $\alpha$, approaching $\eta^{\rm pc}_{\rm CR}\approx 
2\times 10^{-5} (\tau/17\,{\rm Myr})^{1/2} (P/0.253\,{\rm s})^{-3/8}$ 
at $\alpha = 90^{\circ}$ (eq.\ [68] of HM01, after correcting the
numerical coefficients: $0.7\rightarrow 0.97$, $0.05\rightarrow 0.07$,
and $3\rightarrow 0.3$ in the conditions for applicability; 
A.\ Harding 2004, private communication).
This limiting value is a factor of 10 lower than the
$\eta^{\rm pc}$ estimated from the observation, 
so it seems plausible that the PCs of \psr\ 
are heated by CR-produced pairs if $\alpha$ is close to $90^\circ$.

Other models of PC heating have been briefly discussed by HM01.
As a rule, they predict PC luminosities much higher than observed
(see, e.g., Zavlin \& Pavlov 1998),
and their applicability to old, low-$\ed$ pulsars is unclear.

\subsection{X-rays from other old pulsars} 
With the characteristic age of 17 Myrs, \psr\ remains 
the oldest ordinary pulsar detected outside the radio range,
and its spin-down power, $\ed=5.6\times 10^{32}$ erg s$^{-1}$,
is among the lowest in the sample of spin-powered pulsars 
observed in X-rays.  However, there are several 
old ($\tau\gapr 1$ Myr) pulsars, with low spin-down powers 
($\ed \lapr 10^{34}$ erg s$^{-1}$), that have been
observed in X-rays for sufficiently long time to make a useful comparison
with \psr. Six such pulsars, in addition to \psr, 
are presented in Table 3, with the relevant pulsar parameters.
The X-ray properties of three of them have not been previously reported.

The youngest of the objects included in 
Table 3 is PSR B2224+65,
which powers the Guitar nebula (Cordes, Romani, \&  Lundgren 1993). 
The {\sl Chandra}
ACIS\footnote{Advanced CCD Imaging Spectrometer.} observation
in 2000 October nicely resolved the pulsar from the nebula.
We found that the pulsar's X-ray spectrum  
(about 80 counts detected in 50 ks)
can be described by a PL with $\Gamma = 1.73$ and
$L_{\rm 2-10\,keV}^{\rm nonth} = 3.9\times 10^{30}$ erg s$^{-1}$
(the luminosity estimates for this and the other pulsars below
are for the distances given in 
Table 3). Its nonthermal X-ray efficiency, 
$\enon_{\rm 2-10\, keV}=0.0033$, 
exceeds the ``maximum efficiency'' suggested by Possenti et al.\ (2002;
see \S\,5.1) by about 40\%. 
Adding to this best fit a hydrogen PC model with $\alpha=\zeta=45^\circ$,
$(M/M_\odot)/(R/10\,{\rm km})=1.4$, $B=1\times 10^{12}$ G, and
$\rpc = \rpc^*=145\,(P/1\,{\rm s})^{-1/2}$ m
(see \S\,3.2 and \S\, 5.2), 
we can crudely estimate an upper limit on the PC luminosity
(as discussed in \S\,5.2, the inferred $\lbolpc$ is almost
independent of the assumed model of PC emission).
The $ 1\sigma$ upper limit on the PC efficiency of PSR~B2224+65, 
$\epc\sim 4\times 10^{-4}$,
is a factor of 7 lower than the $\epc_{\rm CR}$
predicted by the HM01 model for $\alpha=60^\circ$ 
(as estimated from their Fig.\ 7
by scaling $\epc_{\rm CR}\propto P^{3/2}$, 
according to eq.\ [67] of HM01).

PSR~J2043+27 is of about the same characteristic age
as the pulsar B2224+65, but its spin-down power is a factor of 47 higher. 
Correspondingly, its X-ray properties are quite different.
Our analysis of the \xmm\ observation of 2002 November
(about 100 source counts in a 12 ks exposure)
shows that the pulsar's X-ray emission, detected
only at $E\lapr 2$ keV, is predominantly
thermal (a single PL fit yields a 
photon index $\Gamma\approx 5$, much larger than $\Gamma =1$--2 
for nonthermal X-ray radiation of spin-powered pulsars).
The BB fit yields $\tbb\simeq 0.9$ MK and $\rbb\simeq 2$ km. 
The $\rbb$ looks too large compared with the conventional PC
radius, $\rpc^* = 470$ m, and the bolometric luminosity,
$\lbol^\infty \approx 2\times 10^{31}$ erg s$^{-1}$, is somewhat higher 
than the maximum value expected from the HM01 model.
The NS hydrogen atmosphere models
give $T\simeq 0.5$ MK and a radius of emitting area of about 9 km,
indicating that the detected X-ray emission originates from the
bulk of NS surface. The bolometric surface luminosity in this model is 
$\lbol\approx 4\times 10^{31}$ erg s$^{-1}$.

The surface temperature of PSR~J2043+2740, as yielded by the
NS atmosphere fits, is typical for middle-aged pulsars,
suggesting that its true age may be smaller than the characteristic
age (as one could expect from its short period and high
spin-down power).
Assuming a photon index $\Gamma=1.5$ for a possible nonthermal
component, we put an upper limit on the nonthermal luminosity,
$L_{\rm 2-10\, keV}^{\rm nonth} < 3.8\times 10^{30}$ erg s$^{-1}$, 
such that the X-ray efficiency, $\enon_{\rm 2-10\, keV}
<6.8\times 10^{-5}$, is a factor of 20 lower 
than that of \psr, and it is a factor of 
200 lower than the maximum efficiency for this $\ed$.
Assuming that the detected thermal emission is due to NS cooling,
the upper limit on PC efficiency 
is the lowest among those found for the other pulsars in 
Table 3.
However, this result should be taken with caution, given the small number
of counts detected and strong correlation between the properties
of the two possible thermal components.
A deeper observation is required to constrain the parameters with
better certainty.

The 2.8-Myr-old pulsar B0628--28 is 
the most puzzling object of this group because it was
found to be extremely luminous in the X-ray domain
(\"Ogelman \& Tepedelenlio\v{g}lu 2004).
Our analysis of the 20-ks ACIS data of 2001 November
shows that the source spectrum is best fitted
with a PL model of $\Gamma=2.73$, 
which is surprisingly steep for a spin-powered pulsar.
The estimated X-ray efficiency in the 2--10 keV range is 
about 0.015, much higher than for any other spin-powered pulsar detected
in X-rays, and it is well above 
the maximum efficiency discussed in \S\,5.1.
Single-component thermal models fail to give statistically 
acceptable fits of the observed spectrum 
(e.g., the best BB fit yields $\chi^2_\nu=4.4$ for 9 d.o.f.).
Further analysis of this source, including optical observations, 
is needed to confirm that this is indeed the pulsar counterpart.

Another 3-Myr-old pulsar, B1813--36, 
was not detected in a 30 ks {\sl Chandra}
ACIS exposure of 2001 October (\"Ogelman \& Tepedelenlio\v{g}lu 2004). 
An upper limit on its X-ray efficiency, 
$\enon_{\rm 2-10\, keV} < 4.4\times 10^{-4}$,
is a factor of 2.5 lower than that for \psr.
The upper limit on PC luminosity is very close to that predicted from
the HM01 results.

\psrr\ has about the same age as the two previous pulsars, but it is
much closer. Its X-ray luminosity was estimated from an
{\sl ASCA} observation (Wang \& Halpern 1997).  
Assuming that the detected emission is nonthermal, 
the pulsar's luminosity is
$\lnon_{\rm 2-10\,keV}\simeq 2.1\times 10^{30}$ erg s$^{-1}$,
or $5.5\times 10^{-4}$ of the spin-down power,
for $d=330$ pc derived by Brisken et al.\ (2002) from the
pulsar's parallax (another estimate,
$d\simeq 360$ pc, obtained by Chatterjee et al.\ 2004,
increases the luminosity and efficiency by about 20\%).
The optical radiation of this pulsar 
(Pavlov et al.\ 1996; Mignani et al.\ 2002)
is most likely of a nonthermal origin, with an optical efficiency close
to that of \psr. The same {\sl ASCA} spectrum could be interpreted
as thermal (BB) emission from pulsar's PCs of a temperature
of about 0.5 MK, radius $\rbb\simeq 600$ m, and bolometric luminosity of 
$2.2\times 10^{30}$ erg s$^{-1}$ (for $d=330$ pc).
The corresponding efficiency of one PC, 
$\epc=2.8\times 10^{-4}$, is very close to that of \psr,
and it is a factor of 7 lower than $\epc_{\rm CR}$ estimated
from Figure 7 of HM01. The difference is not as large as for \psr, but we
should not forget that the actual $\epc$ may be lower than
obtained from the BB fit if the contribution of 
the nonthermal component is not negligible.
New data on \psrr\ collected with \xmm\ in 2003 October
and 2004 April
are expected to determine the origin of its X-ray emission and measure
the contributions of the thermal and nonthermal components.

Another nearby old pulsar, B0823+26, 
was observed with \xmm\ in 2002 April,
but the pulsar was only marginally detected, 
at a count rate of $0.003\pm 0.001$ counts s$^{-1}$ in the effective
EPIC-pn exposure of 34 ks.
We conservatively consider this as an upper limit until the detection
is confirmed with another observation.
We estimated an upper limit on the nonthermal X-ray luminosity:
$\lnon_{\rm 2-10\, keV}<6.0\times 10^{28}$ erg s$^{-1}$
for a PL photon index $\Gamma=1.5$.
The corresponding upper limit on the nonthermal efficiency,
$\enon_{\rm 2-10\,keV} < 1.3\times 10^{-4}$, 
is at least a factor of 10 smaller than the efficiency found for \psr,
and a factor of 6 smaller than for \psrr, whose $\ed$ and $\tau$
are very close to those of PSR~B0823+26.
The upper limit on $\epc$ is a factor of 70 lower than the
estimate for $\epc_{\rm CR}$ obtained from Figure 7 of HM01,
but it is 3 orders of magnitude higher than $\epc_{\rm ICS}$
obtained from Figure 8 of HM02.

Thus, the analysis of this sample of old 
pulsars shows quite different properties of X-ray emission. 
The most powerful PSR J2043+2740 manifests
itself as a cooling NS, similar to middle-aged pulsars. 
Some of the pulsars
(J2043+2740, B0823+26) show rather low upper limits on
the nonthermal X-ray efficiency, which means that not all 
old pulsars are as X-ray efficient as
\psr. On the other side,
the X-ray efficiency of PSR B2224+65 is higher than those of \psr\
and \psrr, while PSR~B0628--28 apparently shows an extremely high
efficiency, well above the maximum expected value. 

The nearest pulsars in this sample, 
B0950+08 and B1929+10, show evidence of
PC emission, with bolometric luminosities 
$\lbolpc\sim 3\times 10^{29}$ and $\leq 2\times 10^{30}$ erg s$^{-1}$,
respectively. The upper limits on $\lbolpc$ derived for
the other pulsars 
are in a range of $(1$--$30)\times 10^{29}$ erg s$^{-1}$.
For all the pulsars, except for the most distant (and strongly absorbed)
B1813--36, the estimated efficiencies (or upper limits)
fall below the model predictions of HM01 for $\epc_{\rm CR}$
(B0823+26 being particularly ``underheated''),
but they are well above the predictions of HM02 
for $\epc_{\rm ICS}$. Since most of these pulsars, except for J2043+2740, 
are close to (slightly above) the ``CR death line'',
it hints that the model of HM01 may need a revision for such pulsars. 

\section{Summary}
The \xmm\ observation of the old pulsar \psr\ allowed us to investigate
its timing and spectral properties in X-rays.  The timing
analysis revealed the pulsations of the X-ray flux,
with the pulse shape and pulsed
fraction depending on photon energy. At $E\gapr 0.5$ keV
the observed light curve shows two relatively narrow peaks
with a high pulsed fraction, $\approx 60\%$, which is
a strong indication of a nonthermal (magnetospheric) origin
of the X-ray emission. The light curve
extracted at lower energies shows a broad single
pulse with a lower pulsed fraction.
This energy dependence can be explained by the presence
of another, most likely thermal, component.

The spectral analysis of the detected X-ray radiation 
supports this interpretation. The X-ray spectrum of
\psr\ can be described by a two-component, thermal plus 
nonthermal, model.
We suggest that the thermal component is emitted from two 
hot ($T\approx 1$ MK) spots (PCs) around the pulsar magnetic poles,
covered by a hydrogen atmosphere. Its spectrum peaks
at lower energies, which can explain the energy dependence of 
the X-ray pulse profiles. The PC radius, $\rpc \approx 250$ m, 
estimated from the magnetized NS hydrogen atmosphere models,
is in a good agreement with the predictions of the pulsar models.
This PC model is consistent with the inferred
light curve of the thermal component
at reasonable values of pulsar parameters. The PC luminosity, 
$L_{\rm bol}^{\rm pc}\approx 3\times 10^{29}$ erg s$^{-1}$,
is in agreement with the model by HM01 for PC heating
by positrons generated by curvature radiation 
if the spin and magnetic axes of the pulsar are nearly orthogonal.

The two-component interpretation of the X-ray emission
combined with the fluxes measured in five optical-UV bands implies
that the spectral slope
of the magnetospheric component remains the same, $\Gamma=1.35\pm 0.05$,
in a broad energy range, from 1 eV to 10 keV, 
and the magnetospheric luminosity
and efficiency are about 
$1\times 10^{30}$ erg s$^{-1}$ and $2\times 10^{-3}$ in this energy range.
Both X-ray and optical magnetospheric emission of \psr\
look more luminous than it could be expected from extrapolations of
empirical $L$-$\ed$ correlations
derived for younger, more powerful pulsars. 
However, the X-ray-to-optical luminosity ratio is about the same
as for other pulsars observed in both X-rays and optical.
This implies a close connection between
(possibly, the same mechanism for)
the optical and X-ray magnetospheric emission, while 
the fact that \psr\ is apparently ``overluminous'' in comparison 
with some (but not all) younger pulsars is associated with
geometrical effects (beaming of radiation and orientations of pulsar axes)
that are also responsible for the large scatter 
in the $L_X$-$\ed$ correlation.

The comparison of \psr\ with six other old pulsars observed in X-rays
reveals very diverse properties of the X-ray spectra and luminosities
of old pulsars. In particular, the estimated magnetospheric 
X-ray efficiencies of these pulsars 
vary from $<10^{-4}$ to $\sim 10^{-1}$. 
The large scatter in a narrow range of characteristic
ages (or spin-down powers) confirms that some ``hidden'' parameters
(e.g., orientations of pulsar's axes) affect 
the apparent X-ray luminosities of pulsars.
Most of the PC luminosities (or their upper limits) estimated
for this group of pulsars fall below the model predictions 
for PC heating by relativistic particles produced
through curvature radiation,
indicating that the models may require a revision for pulsars
near the ``death line'' for this pair production mechanism.

Finally, the derived upper limit of 0.15 MK on the NS brightness 
temperature in the optical-UV range (for a standard NS radius) 
constrains heating mechanisms
possibly operating in the NS interiors.
The actual temperature (or a more restrictive upper limit) 
can be measured from far-UV observations of \psr.

\acknowledgements
We thank Wojciech Lewandowski for providing 
the radio data for \psr,
and Alice Harding, Oleg Kargaltsev, Christian Motch, 
Allyn Tennant, and Sergei Zharikov for useful discussions.
The work of GGP was supported by NASA grant NAG5-10865.


\clearpage

\voffset=-0.4truein
\begin{deluxetable}{ll}
\tabletypesize{\scriptsize}
\tablewidth{0pt}
\tablecaption{
Observed parameters for the radio pulsar B0950+08} 
\tablehead{\colhead{Parameter} & \colhead{Value} } 
\startdata
R.A.~(J2000)..............................................
& ~~~~~~~~$09^{\rm h}53^{\rm m}09\fs307(1)$~~~~~~ \\
Dec.~(J2000)..............................................
& ~~~~~~~~$+07^\circ 55' 36\farcs15(2)$~~~~~~~~~~~~ \\
Epoch of position (MJD)...........................
& ~~~~~~~~51,544.0 \\
Proper motion 
in R.A./Dec. (mas yr$^{-1}$)...
& ~~~~~~~~$-$2.09(8)/29.46(7) \\
Spin frequency, $f$~(Hz)...............................
&  ~~~~~~~~3.951549502802(9) \\
Frequency derivative, $\dot{f}$~(10$^{-15}$~s$^{-2}$)...........
& ~~~~~~~~$-3.596(2)$~~~~~~~~~~~~~         \\
Epoch of frequency (MJD)........................
& ~~~~~~~~52,403.0~~~~~~~~~~~~~~~~~~~~   \\
Timing data span (MJD)...........................
& ~~~~~~~~52,280--52,630 \\
Dispersion Measure, DM (cm$^{-3}$~pc)\,..........
& ~~~~~~~~2.958(3) \\
Epoch of radio pulse (MJD~TDB).............
& ~~~~~~~~52,402.848253347(2) \\
Distance, $d$ (pc)..........................................
& ~~~~~~~~262(5) \\
\enddata
\tablecomments{Based on radio observations made 
at the Torun Radio Astronomy Observatory.
The pulsar position, proper motion, and distance are from
Brisken et al.\ (2002).
The numbers in parentheses represent uncertainties in 
the least significant digit. 
}
\end{deluxetable}
\clearpage
\newpage

\begin{deluxetable}{lllll}
\tabletypesize{\scriptsize}
\tablewidth{0pt}
\tablecaption{Pulsar luminosities}
\tablehead{\colhead{PSR} & \colhead{$d$} & 
\colhead{$\log \ed$} & \colhead{$\log L_{\rm opt}^{\rm nonth}$} &
\colhead{$\log\lnon_{X}$}  \\
 & \colhead{(kpc)} & \colhead{~(erg\,s$^{-1}$)} &
\colhead{(erg\,s$^{-1}$)} & \colhead{(erg\,s$^{-1}$)}  }
\startdata
Crab & 2.0 & ~~~38.65 & $33.80\pm0.05$ & $36.18\pm0.05$  \\
B0540--69 & 50.0 &  ~~~38.17 & $33.94\pm0.10$ & $36.38\pm0.07$  \\
Vela  & 0.29 & ~~~36.84 & $28.80\pm0.07$ & $31.30\pm0.04$  \\
B0656+14 & 0.29 & ~~~34.58 & $28.20\pm0.12$ & $30.15\pm0.08$  \\
Geminga & 0.20 & ~~~34.51 & $27.49\pm0.10$ & $30.19\pm0.06$  \\
B1929+10\tablenotemark{a} & 0.33 & ~~~33.59 & $27.61\pm0.12$ & $30.58\pm0.08$ \\
B0950+08 & 0.26 & ~~~32.75 & $27.36\pm0.10$ & $29.92\pm0.06$  \\
\enddata
\tablecomments{
Fourth and fifth columns give the nonthermal optical
(4000--9000 \AA) and X-ray (1--10 keV)
luminosities computed for the distances given in
second column. 
The X-ray luminosities of the Crab and Vela pulsars are measured from
the {\sl Chandra} observations of 2000 February 2 and 2003 August 20,
respectively. For B0950+08,
$\lnon_{X}$ is the luminosity of
the PL component in the PL+PC model (\S3.2),
while the luminosities of the other four pulsars
are taken from the works of
Kaaret et al.\ (2001; B0540--69), Pavlov et al.\ (2002; B0656+14),
Zavlin \& Pavlov (2004; B0656+14 and Geminga),
and Wang \& Halpern (1997; B1929+10).
The optical luminosities are estimated from power-law fits of
the nonthermal spectral components
(a power-law extrapolation of the UV spectrum in the case of B1929+10),
using the works of Sollerman et al.\ (2000; Crab),
Middleditch et al.\ (1987; B0540--69), Shibanov et al.\ (2003; Vela),
Koptsevich et al.\ (2001; B0656+14 and Geminga),
Mignani et al.\ (2002; B1929+10), and Zharikov et al.\ (2004; B0950+08),
as well as recent {\sl HST} observations
of Vela, B0656+14, and Geminga.
The errors for the $L_{\rm opt}^{\rm nonth}$ values
include our assessment of systematic uncertainties.
}
\tablenotetext{a}{
Since it is assumed that the X-ray emission
of B1929+10 is fully nonthermal while it may have a PC component,
the corresponding optical-to-X-ray
luminosity ratio can be considered as a lower limit.
}
\end{deluxetable}
\clearpage
\newpage


\voffset=-0.4truein
\begin{deluxetable}{llllllllll}
\tabletypesize{\scriptsize} 
\tablewidth{0pt} 
\tablecaption{Properties of old pulsars observed in X-rays}
\tablehead{\colhead{PSR} & \colhead{$P$} & \colhead{$\tau$} & 
\colhead{$\log\ed$} & \colhead{$d$} & \colhead{$\log F_X$} &
\colhead{$\log\lnon$} & \colhead{$\log\enon$} &
\colhead{$\log\lbolpc$} & \colhead{$\log\epc$} \\
& \colhead{(s)} &  \colhead{(Myr)}  & \colhead{(erg\,s$^{-1}$)} 
& \colhead{(kpc)} & \colhead{(erg\,s$^{-1}$\,cm$^{-2}$)} &
\colhead{(erg\,s$^{-1}$)} &    & \colhead{(erg\,s$^{-1}$)} & }
\startdata
B2224+65  & 0.683 & ~\,1.1 & \,~33.08 & \,1.86 & ~~\,\,$-13.85$ &
 ~~\,\,30.87 & ~~\,\,$-2.21$ & $<29.96$ & $<-3.42$ \\
J2043+2740\tablenotemark{a} & 0.096 & ~\,1.2 & \,~34.75 & \,1.80 &
~~\,\,$-13.96$ & $<30.75$ & $<-4.00$ & $<30.36$ & $<-4.69$ \\
B0628--28 & 1.244 & ~\,2.8 & \,~32.45 & \,1.45 &
~~\,\,$-13.27$ & ~~\,\,31.52 & ~~\,\,$-0.93$ & $<30.18$ & $<-2.57$ \\
B1813--36 & 0.387 & ~\,3.0 & \,~33.15 & \,2.44 &
$<-14.82$ & $<30.18$ & $<-2.97$ & $<30.44$ & $<-3.01$ \\
B1929+10\tablenotemark{b} & 0.227 & ~\,3.1 & \,~33.59 & \,0.33 &
~~\,\,$-12.39$ & ~~\,\,30.74 & ~~\,\,$-2.85$ &  
~~\,\,30.34 & ~~\,\,$-3.55$ \\
B0823+26 & 0.531 & ~\,4.9 & \,~32.65 & \,0.34 &
$<-14.16$ & $<29.00$ & $<-3.65$ & $<29.13$ & $<-3.82$ \\
B0950+08 & 0.253 & 17.4 & \,~32.75 & \,0.26 &
~~\,\,$-12.96$ & ~~\,\,30.01 & ~~\,\,$-2.74$ & 
~~\,\,29.43 & ~~\,\,$-3.62$ \\
\enddata
\tablecomments{
Second through fourth columns give standard radio pulsar parameters.
The observed (absorbed) fluxes, $F_X$,
nonthermal (isotropic) luminosity, $\lnon$, and 
nonthermal X-ray efficiency, $\enon=\lnon/\ed$,
are for the 0.2--10 keV band.
The bolometric PC efficiency is
defined as $\epc=\lbolpc/(2\,\ed)$.
The X-ray parameters of PSRs B1813--36 and B1929+10 are from
\"Ogelman \& Tepedelenlio\v{g}lu (2004) and Wang \& Halpern (1997),
respectively, whereas those for the other pulsars are from this work.
The upper limits on $\lnon$ 
are derived with a PL model of photon index $\Gamma=1.5$.
The upper limits on $\lbolpc$ are obtained
assuming the conventional PC radius
$\rpc^*= 145\, (P/1\,{\rm s})^{-1/2}$ m (see \S\,5.2).
The distances to PSRs B1929+10 and 0950+08 are inferred from
the pulsars' parallaxes (Brisken et al.\ 2002). The other
distances are estimated from the pulsar dispersion measures
and the model of Galactic distribution of free electrons
(Cordes \& Lazio 2003).
}
\tablenotetext{a}{
The upper limits for PSR~J2043+2740 are obtained 
assuming that the detected thermal X-ray radiation 
is emitted from the bulk of the NS surface (see \S\,5.3).
}
\tablenotetext{b}{
The luminosity estimates for \psrr\ are obtained assuming
that all the detected X-ray radiation is either nonthermal or thermal.
}
\end{deluxetable}
\clearpage


\begin{figure}
\plotone{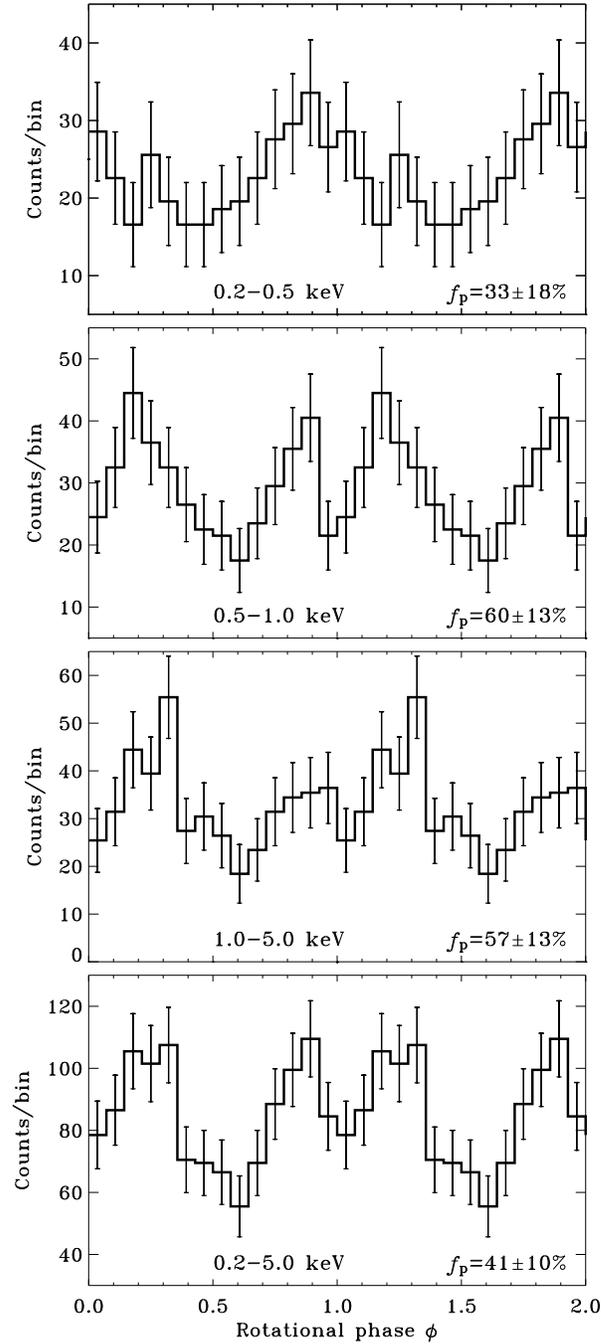}
\vskip -1cm
\caption{X-ray light curves of \psr\ extracted
from the EPIC-pn data in four
energy bands, with the values of the intrinsic pulsed
fraction $\fp$ and its $1 \sigma$ errors. 
The zero phase corresponds to the maximum of radio pulse.
}
\end{figure}
\clearpage

\begin{figure}
\plotone{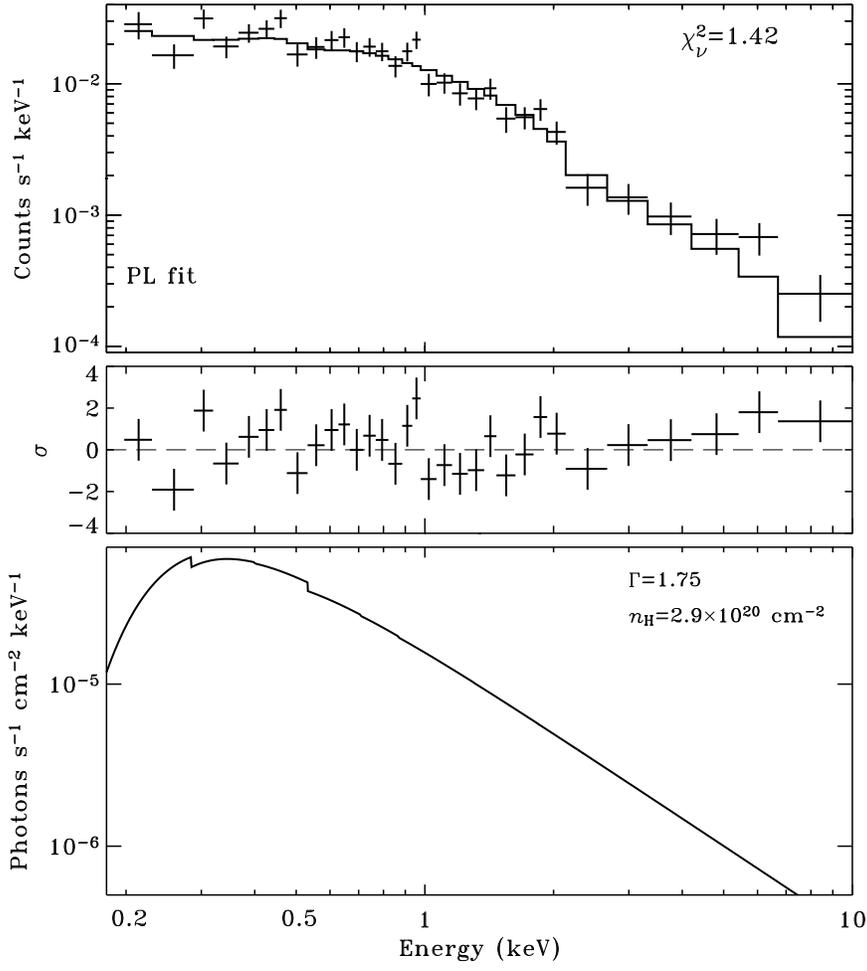}
\vskip -1cm
\caption{Observed EPIC-pn spectrum of \psr\ fitted with
a single power-law (PL) model of photon index $\Gamma=1.75$
and residuals of the fit (two upper panels).
The model spectrum is shown in the lower panel.
}
\end{figure}
\clearpage

\begin{figure}
\plotone{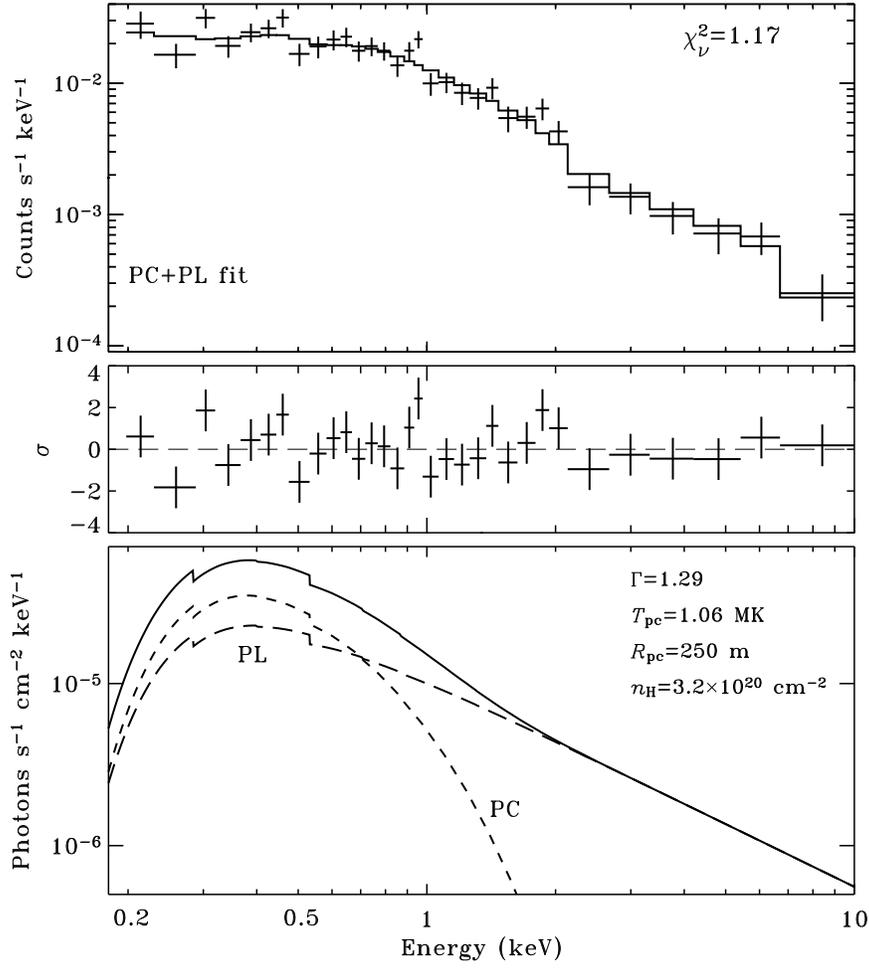}
\caption{Polar-cap (PC) plus power-law (PL) fit to the
observed EPIC-pn spectrum of \psr\ and its residuals (two
upper panels). The spectral components are shown in the lower panel.
See \S\,3.2 for the other PC parameters.
}
\end{figure}
\clearpage

\begin{figure}
\plotone{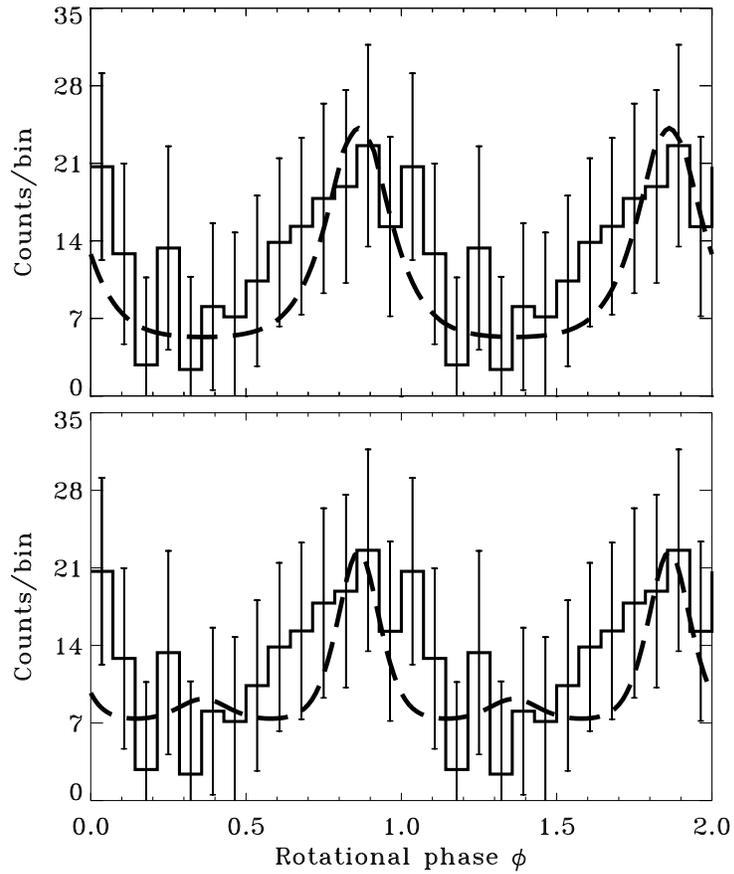}
\vskip -2cm
\caption{
Residual pulse profile (histograms) in the 0.2--0.5 keV 
band after subtracting the scaled 1--5 keV light curve. 
Thick dashes show the pulse profiles for the PC models with
the angles $\zeta=35^\circ$, $\alpha=30^\circ$ (upper
panel), and $\zeta=53^\circ$, $\alpha=75^\circ$ (lower
panel). See \S\,3.3 for the other PC parameters.
}
\end{figure}
\clearpage

\begin{figure}
\plotone{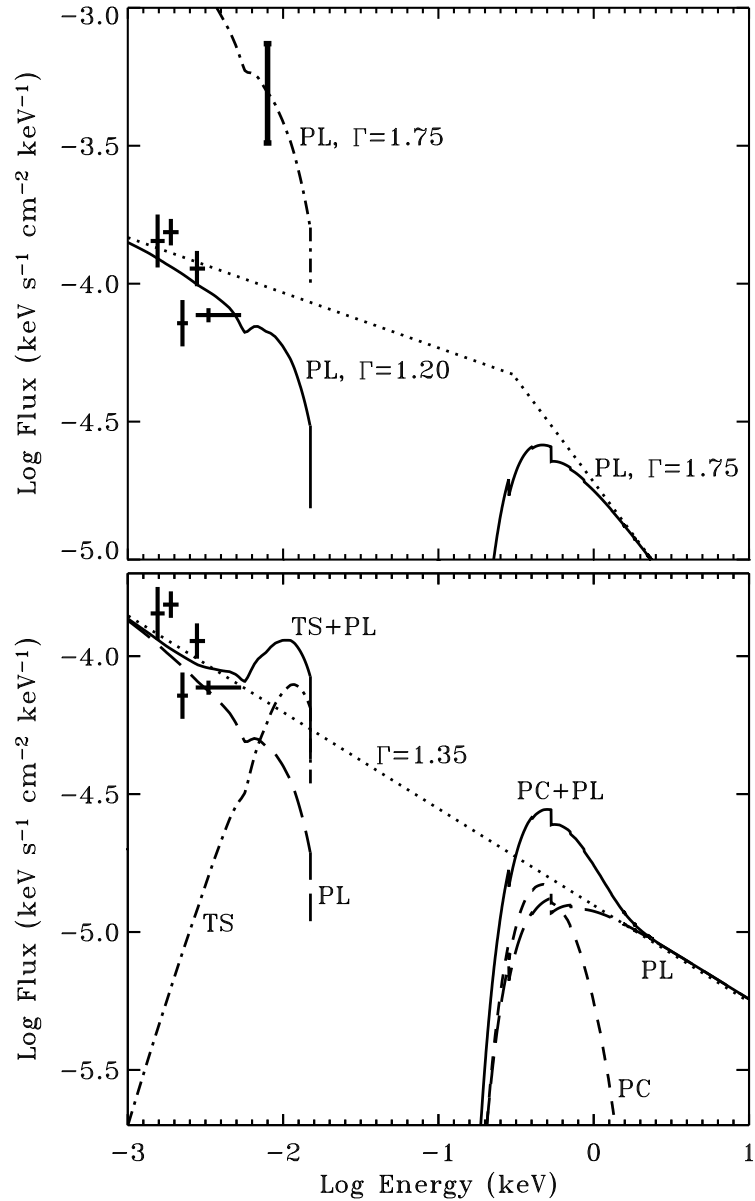}
\vskip -2cm
\caption{
Broad-band (optical through X-ray) spectrum of \psr\ for two  models
of its X-ray spectrum: single PL (upper panel) and PC+PL (lower panel).
The model spectra are shown by the solid lines in the X-ray range;
the PC model is the same as used for modeling the pulse profile in the
upper panel of Fig.\ 4. Crosses show the observed fluxes in 
optical-UV bands (Pavlov et al.\ 1996; Zharikov et al.\ 2004). 
The optical spectrum is modeled by an absorbed PL
with $\Gamma=1.20$ and 1.35 
(solid and dashed lines in the upper and lower panels, 
respectively), and $E(B-V)=0.05$.
The dash-dot line in the upper panel shows an extrapolation of 
the X-ray PL spectrum into the optical; 
the thick vertical bar demonstrates the propagated errors.
The dotted lines are the unabsorbed spectra of the nonthermal components:
single and broken PLs in the lower and upper panels, respectively. 
The solid line in the optical part of the lower panel is the sum of the
PL component and a possible thermal spectrum (TS) from
the whole NS surface, for a 0.1 MK temperature
(see \S\,4 for other model parameters).
}
\end{figure}
\clearpage

\begin{figure}
\plotone{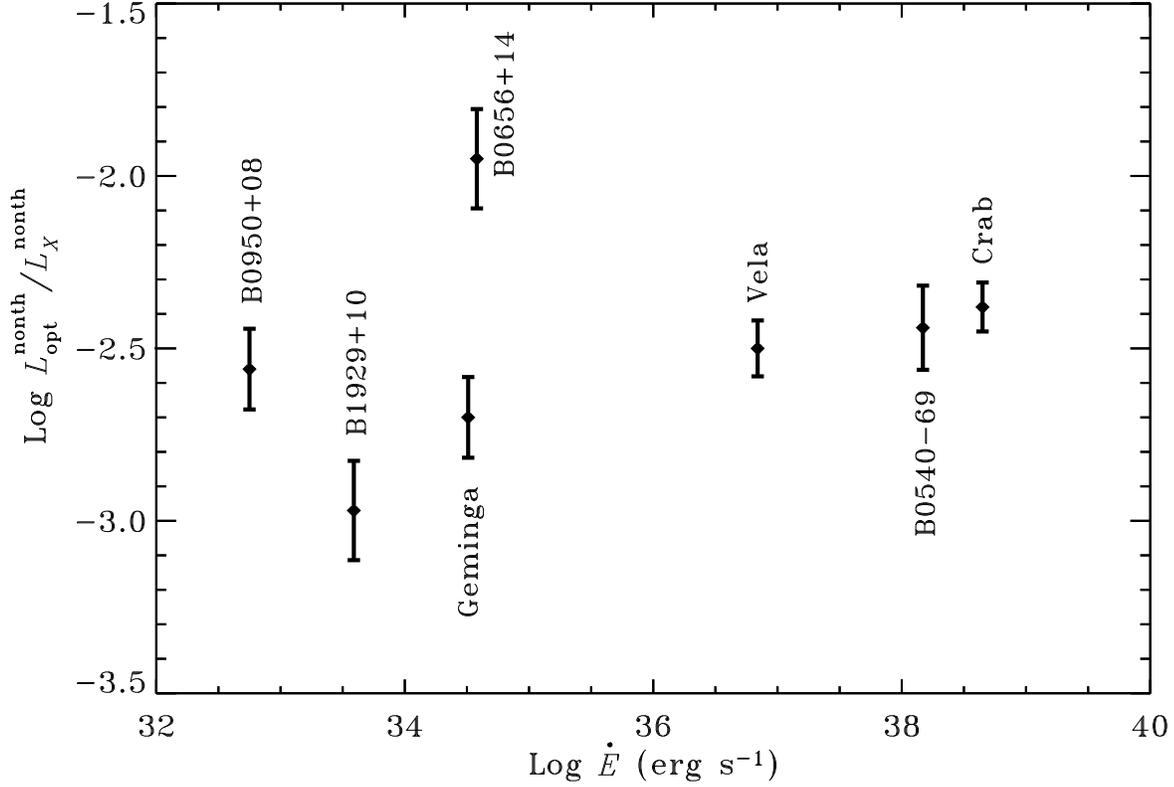}
\vskip -4cm
\caption{
Ratio of the nonthermal optical luminosity in the 4000--9000 \AA\
band to the nonthermal component of the X-ray luminosity in the 1--10
keV band versus spin-down power $\ed$ for seven pulsars.
See text and Table 2 for more details.
}
\end{figure}
\clearpage

\end{document}